\documentclass[12pt]{article}
\usepackage[margin=1in]{geometry}
\usepackage{setspace}
\usepackage{natbib}
\usepackage{newfloat}
\usepackage{amsmath}
\usepackage{amsfonts}
\usepackage{amssymb}
\usepackage{lineno}
\usepackage{parskip}
\DeclareFloatingEnvironment[name={Extended Data Figure}]{suppfigure}
\usepackage[english]{babel}

\usepackage{graphicx}
\usepackage{xcolor}

\definecolor{darkgreen}{rgb}{0.0, 0.5, 0.0}
\date{}

\begin{document}

\title{Manufacturing Complex Structures Through Interfacial Self-Organization}

\author{
Jacob Garrett$^{1}$, Abhinit Sati$^{2}$, Edward Chien$^{2}$, Emily Whiting$^{2}$,\\ and Valeri Frumkin$^{1}$\\
$^{1}$Department of Mechanical Engineering, Boston University, Boston, MA, USA\\
$^{2}$Department of Computer Science, Boston University, Boston, MA, USA
}

\maketitle

\begin{abstract}
Conventional manufacturing constructs three-dimensional geometries through localized material deposition, removal, or shaping, with fabrication effort greatly increasing with geometric complexity. Here, we introduce a manufacturing approach in which target geometry is encoded into sparse wireframe scaffolds that guide liquid self-organization through interfacial forces. We apply a computational framework that approximates arbitrary target geometries using constant-mean-curvature surface domains and extracts their boundaries to construct the scaffolds. Under neutral buoyancy conditions, curable liquids injected into these scaffolds self-organize into the prescribed geometries, which are subsequently solidified into mechanically isotropic structures with optically smooth surfaces. We demonstrate the fabrication of complex anatomical models and topology-optimized structures featuring intricate branching geometries, internal cavities, and integrated materials. This approach opens new possibilities for scalable fabrication of complex geometric designs.
\end{abstract}

\subsection*{Introduction}\label{sec1}

Advances in computational design have generated structures of remarkable efficiency and performance. Algorithmic design tools such as Topology Optimization (TO) \cite{bendsoe_topology_2004, sigmund_topology_2013} can be used to optimize the distribution of material within computational models to meet desired objectives such as material efficiency \cite{berrocal_topology_2019, tang_topology_2024}, mechanical response  \cite{garland_pragmatic_2021, eschenauer_topology_2001}, thermal management \cite{lin_multi-objective_2025, dbouk_review_2017, bruns_topology_2007}, and aerodynamic performance \cite{gomes_aerodynamic-driven_2020, james_concurrent_2014}.
Significant improvements to these metrics make TO designs highly desirable across multiple industries, including robotics \cite{sha_topology_2020, wang_optimal_2019, yap_topology_2023}, aerospace \cite{berrocal_topology_2019, zhu_topology_2016}, vehicle design \cite{jankovics_customization_2019, walton_design_2017}, and medical prosthetics \cite{iqbal_general_2019, jansari_comparative_2019}.
However, the resulting designs frequently incorporate complex geometric features such as overhangs, undercuts, highly curved surfaces, and enclosed cavities \cite{liu_survey_2024, stankiewicz_geometrically_2022}, that are challenging to realize at scale with conventional fabrication methods. 

Subtractive manufacturing (SM), which encompasses material removal processes such as milling and turning, can reliably make parts from a wide variety of materials with tight dimensional tolerances \cite{groover_fundamentals_2019}. While many SM processes are automated and used in mass production \cite{groover_fundamentals_2019, rahman_review_2023}, they are wasteful \cite{paris_comparative_2016, mohd_yusuf_review_2019}, and often limited by tooling constraints that make the fabrication of parts containing complex features, such as internal voids and undercuts, challenging \cite{boothroyd_product_2010, kerbrat_manufacturing_2010}.
Formative manufacturing (FM) processes such as casting and injection molding are highly scalable due to the reusability of their molds \cite{groover_fundamentals_2019}. Some FM methods also allow for extensive customization of the materials involved, using additives to create composites with desirable properties \cite{dobrota_functional_2025, sarmah_recent_2024}. However, parts with complex topologies may require expensive multi-stage molds or sacrificial molds that cannot be recovered after demolding, when fabrication is feasible at all \cite{osswald_polymer_2006, beaumont_runner_2019}. Furthermore, parts made using FM methods often have loose tolerances \cite{deboer_additive_2021} and require careful temperature management to prevent poor solidification \cite{ahn_evaluation_2025, czepiel_advanced_2023}. 

The most widely used approach for fabricating parts with complex geometric features is Additive Manufacturing (AM), which encompasses a wide variety of processes such as Fused Filament Fabrication \cite{singh_current_2020, kumar_fused_2023}, Stereolithography \cite{melchels_review_2010, huang_review_2020}, and Selective Laser Sintering \cite{kruth_lasers_2003, kruth_binding_2005}. 
While excellent for rapid prototyping and low-unit production, AM has seen limited industrial uptake due to a variety of limitations \cite{tofail_additive_2018}. The discrete layer-by-layer deposition of material, common to most AM processes, results in disadvantageous anisotropic mechanical properties \cite{liu_survey_2024, ahn_anisotropic_2002}, and poor surface roughness \cite{ali_investigation_2023}. Furthermore, fabrication time scales very unfavorably with print volume and resolution, as the deposition or solidification mechanism must visit every voxel of the original model \cite{tofail_additive_2018, gao_status_2015}.
Volumetric AM techniques have recently emerged, where the material is rapidly cured as one continuous volume \cite{tumbleston_continuous_2015, kelly_volumetric_2019}. However, these techniques are highly restricted to a select group of materials and remain constrained in achievable part size \cite{whyte_volumetric_2024}. 

To account for the limitations of existing manufacturing techniques, modified TO algorithms have been developed \cite{harzheim_review_2006, ibhadode_topology_2023}, however, they can severely inhibit the performance gains that TO parts can deliver \cite{dangal_impact_2023, vatanabe_topology_2016}. Therefore, there is a need for alternative manufacturing methods that can overcome these geometric limitations without incurring additional tooling requirements or fabrication time penalties as geometric complexity increases. 

\subsection*{Manufacturing through interfacial self-organization}\label{sec2}
{\it Fluidic Shaping} is a manufacturing approach recently developed for the fabrication of optical components \cite{frumkin_fluidic_2021, elgarisi_fabrication_2021}. This approach leverages interfacial self-organization of curable liquids (e.g., polymers) under suppressed-gravity conditions achieved within a neutral buoyancy fabrication environment. 
The deposited material cures as one continuous bulk, allowing one to fabricate decimeter-scale optical components in minutes \cite{elgarisi_fabrication_2025}.
Notably, due to the natural smoothness of liquid interfaces, fabricated components exhibit exceptional surface roughness, on the order of 1 nm RMS \cite{frumkin_fluidic_2021}.
Recent work has demonstrated how interfacial self-organization can be leveraged to create other shapes as well, such as soft cells constructed using minimal surfaces \cite{domokos_soft_2024} and periodic constant-mean-curvature tilings \cite{hochman_liquifab_2025}.
Here we demonstrate how the {\it Fluidic Shaping} approach can be applied for the fabrication of parts with highly complex geometric features, producing large, smooth and highly isotropic final products from a variety of materials.

\begin{figure}[t!]
\noindent \begin{centering}
\hspace*{-0.5cm}
\includegraphics[width=1\textwidth]{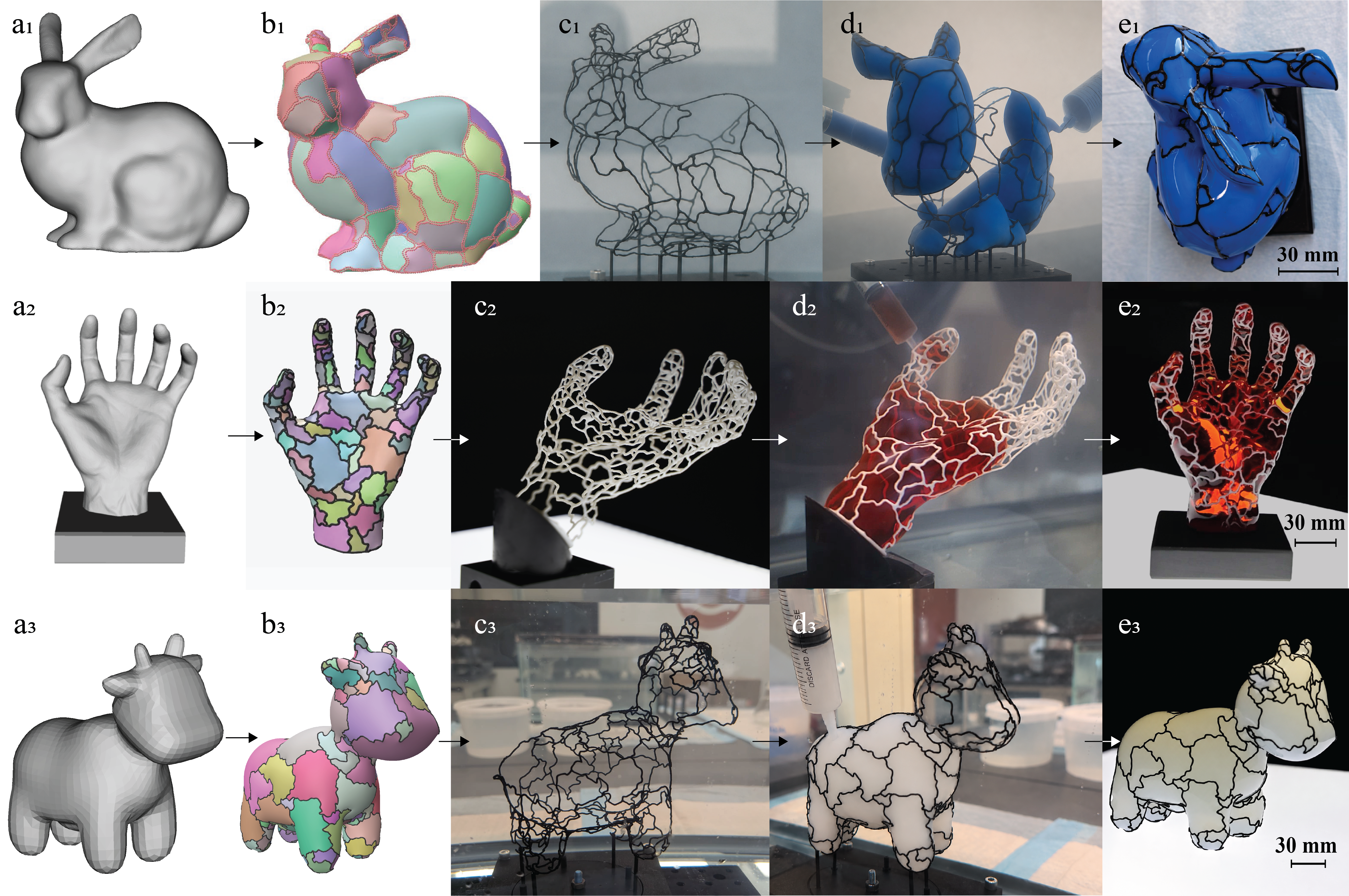}
\par\end{centering}
    \caption{\small{{\bf Fabrication of complex geometries via interfacial self-organization.} Panels a1-3 and b1-3: A three-dimensional target geometry is first approximated by a collection of constant-mean-curvature surface patches. Panels c1-3: The boundaries between adjacent patches define a sparse wireframe that encodes the target geometry and can be fabricated using conventional techniques, e.g., 3D printing, wire bending, and laser cutting. Panels d1-3: The wireframe is submerged in a density-matched immersion liquid simulating microgravity conditions, and a liquid polymer is injected into the frame. Capillary forces drive the polymer interfaces to self-organize into the geometry prescribed by the wireframe. Panels e1-3: Once equilibrated, the polymer is cured, for example thermally or by UV exposure, producing the final smooth, solid structure. The process is demonstrated for three geometrically distinct targets: the Stanford Bunny, an anatomically accurate human hand \cite{noauthor_wallmount_nodate}, and a 3D model of Spot the cow \cite{crane2013robust}. The wireframe in all three examples is made of PLA and PLA Tough+ (Bambu Labs). The Stanford Bunny is made of Mold Star 30 (Smooth-On), the hand model from Vida Rosa TJ-3704A, and Spot the cow from Ecoflex 00-10 (Smooth-On).} }
    \label{fig:1}
\end{figure}

Figure 1 describes the workflow of fluidic fabrication through interfacial self-organization for three different target models. The target geometry is approximated via a collection of constant mean curvature (CMC) domains (Fig. 1 a,b), as described in subsequent sections. A wireframe scaffold is then extracted from the boundaries between these domains (Fig. 1 c), and placed inside an immersion liquid environment consisting of a water-glycerol mixture, the density of which is matched to that of the shaping material (Materials and Methods). A material of choice is then injected into the wireframe, where it self-organizes into a configuration corresponding to its minimum energy state, under the constraints imposed by the geometry of the wireframe and the injected volume (Fig. 1 d). Once the target geometry is obtained, the material can be solidified (e.g., UV cured or thermoset) and removed from the immersion liquid (Fig. 1 e). A detailed demonstration of {\it Fluidic Shaping}-based fabrication is provided in the supplementary video.

The geometry generated through \textit{Fluidic Shaping} arises from the equilibrium configuration of liquid interfaces when gravitational effects can be neglected.  
Under these conditions, the dominant contribution to the system's free energy becomes the interfacial energy associated with surface tension. Therefore, a liquid-fluid interface bounded via wetting by a simple closed curve will evolve spontaneously toward a configuration that minimizes its total surface energy while satisfying the imposed boundary conditions and volume constraints.
The pressure difference across the interface is governed by the Young-Laplace equation
\begin{equation}
\Delta P =2\gamma \kappa
\end{equation}
where $\gamma$ is the interfacial tension and $\kappa$ is the mean curvature of the interface $h(x,y)$, given in its local frame of reference by
\begin{equation}
    \kappa
    = \frac{(1 + h_y^2)\, h_{xx} - 2h_x h_y\, h_{xy}
            + (1 + h_x^2)\, h_{yy}}
           {2\left(1 + h_x^2 + h_y^2\right)^{3/2}}
\end{equation}


(see Supplementary Information). Since at equilibrium, the pressure within a connected liquid domain is spatially uniform, the mean curvature must likewise remain constant throughout the interface, leaving the surface topography to be determined by the geometry of its boundary and the total volume of liquid contained within it. 

\subsection*{CMC decomposition and wireframe extraction}
Since arbitrary 3D shapes cannot generally be represented as CMC surfaces, we approximate their geometry by segmenting their surface into a collection of enclosed patches, with each patch being of a prescribed mean curvature $\kappa_0$. 
To this end, we apply a computational framework that combines surface segmentation with iterative curvature-driven relaxation procedures that evolve individual surface patches toward CMC states while preserving their shared boundaries. 
Once the CMC decomposition is obtained, the boundaries separating neighboring domains are extracted to form a wireframe scaffold suitable for \textit{Fluidic Shaping}-based fabrication. In this manner, the decomposition process establishes a direct mapping between an arbitrary target geometry and a manufacturable set of fluidic boundary conditions. A detailed treatment of the CMC decomposition problem is available in an accompanying publication \cite{sati_piecewise_2026}.

\begin{figure}[t!]
\noindent \begin{centering}
\hspace*{-0.5cm}
\includegraphics[width=1\textwidth]{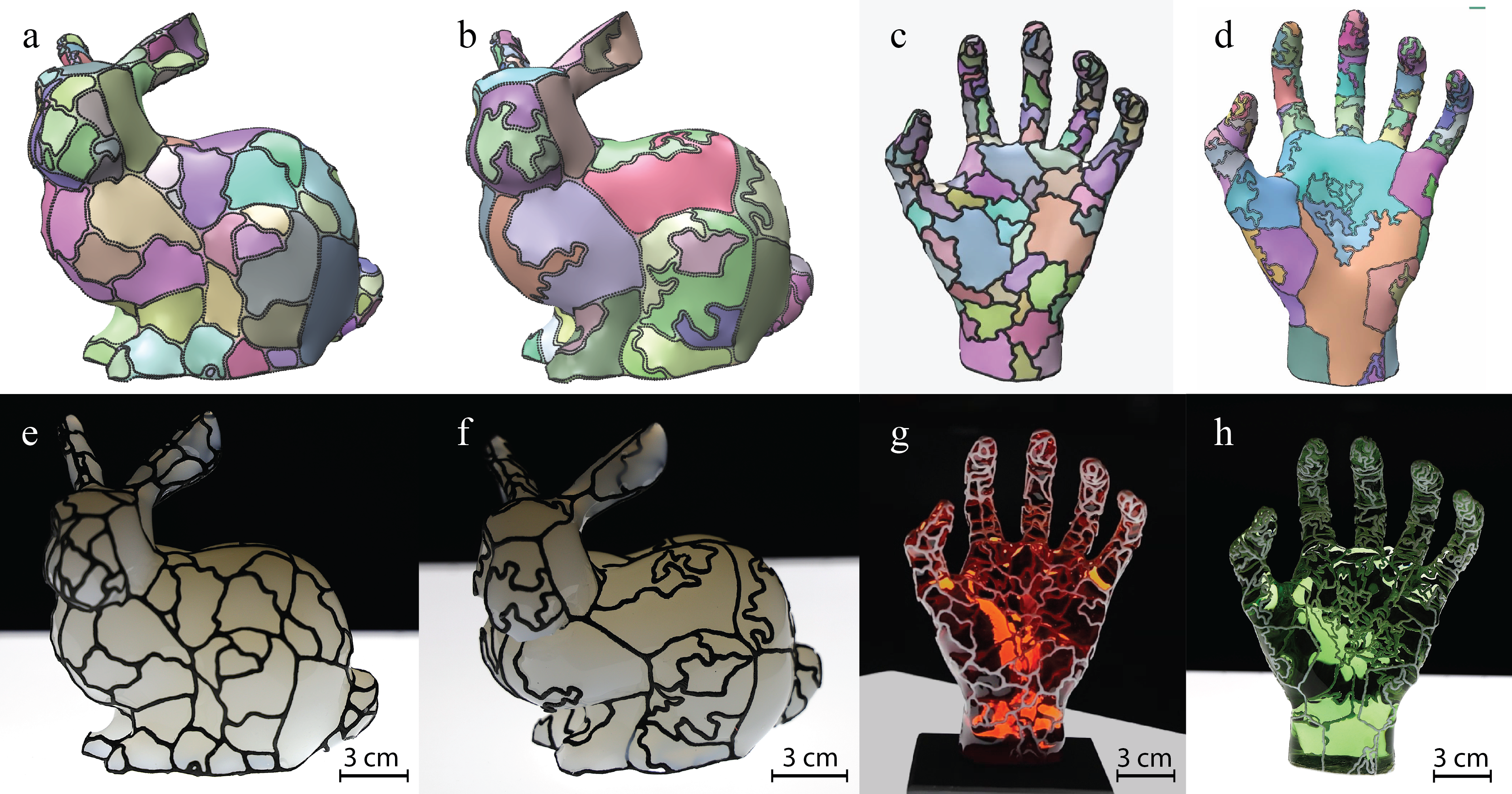}
\par\end{centering}
    \caption{\small{ Non-unique decomposition of a complex geometry into coupled constant-mean-curvature domains. (a) High-compactness decomposition of the Stanford bunny geometry into coupled CMC domains. (b) Lower-compactness decomposition of the same geometry. (c)-(d) High and low-compactness decomposition, respectively, of an anatomically accurate human hand \cite{noauthor_wallmount_nodate}.
    (e)-(h) Corresponding fabricated objects. Although two decompositions may approximate the same target geometry, they produce substantially different domain organizations and scaffold architectures.} }
    \label{fig:2}
\end{figure}

A notable feature of this representation is its inherent non-uniqueness. A given target geometry may admit multiple valid decompositions, each producing a distinct wireframe architecture and corresponding fluidic realization. These alternative representations may differ substantially in geometric fidelity and complexity of the wireframe. Consequently, the decomposition process itself becomes an inverse design problem in which a target geometry may be represented through many different physically realizable wireframes, each optimized for specific requirements. For this goal, we define a compactness measure designed to quantify the level of complexity and manufacturability of a given wireframe.

{\bf Definition:} Let a patch be a triangle mesh $\mathcal{M} = (V, F)$ with vertex
positions $\mathbf{x}_i \in \mathbb{R}^3$. 
The boundary length and surface area of this patch are given by
\[
  L = \sum_{\{i,j\} \in \partial\mathcal{M}} \lVert \mathbf{x}_j - \mathbf{x}_i \rVert,
  \qquad
  A = \sum_{(i,j,k) \in F} \tfrac{1}{2} \bigl\lVert (\mathbf{x}_j - \mathbf{x}_i) \times (\mathbf{x}_k - \mathbf{x}_i) \bigr\rVert.
\]
We then define compactness as the isoperimetric quotient
\[
  C(\mathcal{M}) = \frac{4\pi A}{L^2}.
\]
For planar patches, $C$ is maximized at $1$ for a disk patch, while any other boundary shape results in a lower compactness measure, tending to $0$ for very elongated or jagged patches. This same ratio may exceed $1$ in the case of non-planar patches, but remains a good practical way to avoid overly complex patches, e.g., as used in \cite{garland_hierarchical_2001}.

Figure \ref{fig:2} illustrates this concept using two distinct decompositions of the Stanford bunny geometry and an anatomically accurate model of a human hand \cite{noauthor_wallmount_nodate}. The decompositions shown in Fig. \ref{fig:2}a and \ref{fig:2}c were generated using a high-compactness objective, resulting in a relatively simple wireframe architecture composed of simple CMC domains and boundary segments. In contrast, the decompositions shown in Fig. \ref{fig:2}b and \ref{fig:2}d employ a lower compactness constraint, producing more complex wireframes that are capable of capturing a similar level of geometric detail while providing a substantially larger exposed surface area. Figures \ref{fig:2}e-h show physical realizations of the corresponding wireframe designs, using PLA (Bambu Labs) for the wireframe material, a soft elastomer (Smooth-On EcoFlex 00-30) for the bulk of the Stanford bunny, and a hard UV resin (Vida Rosa TJ-3704A) for the bulk of the hand (see Materials and Methods). Although each pair of decompositions approximate the same target geometry, they generate substantially different scaffold architectures, each with its unique characteristics. Further discussion of the compactness measure can be found in the SI and in \cite{sati_piecewise_2026}.

\subsection*{Interfacial fabrication of topology optimized designs}\label{sec3}

\begin{figure}[t!]
\noindent \begin{centering}
\hspace*{-0.5cm}
\includegraphics[width=1\textwidth]{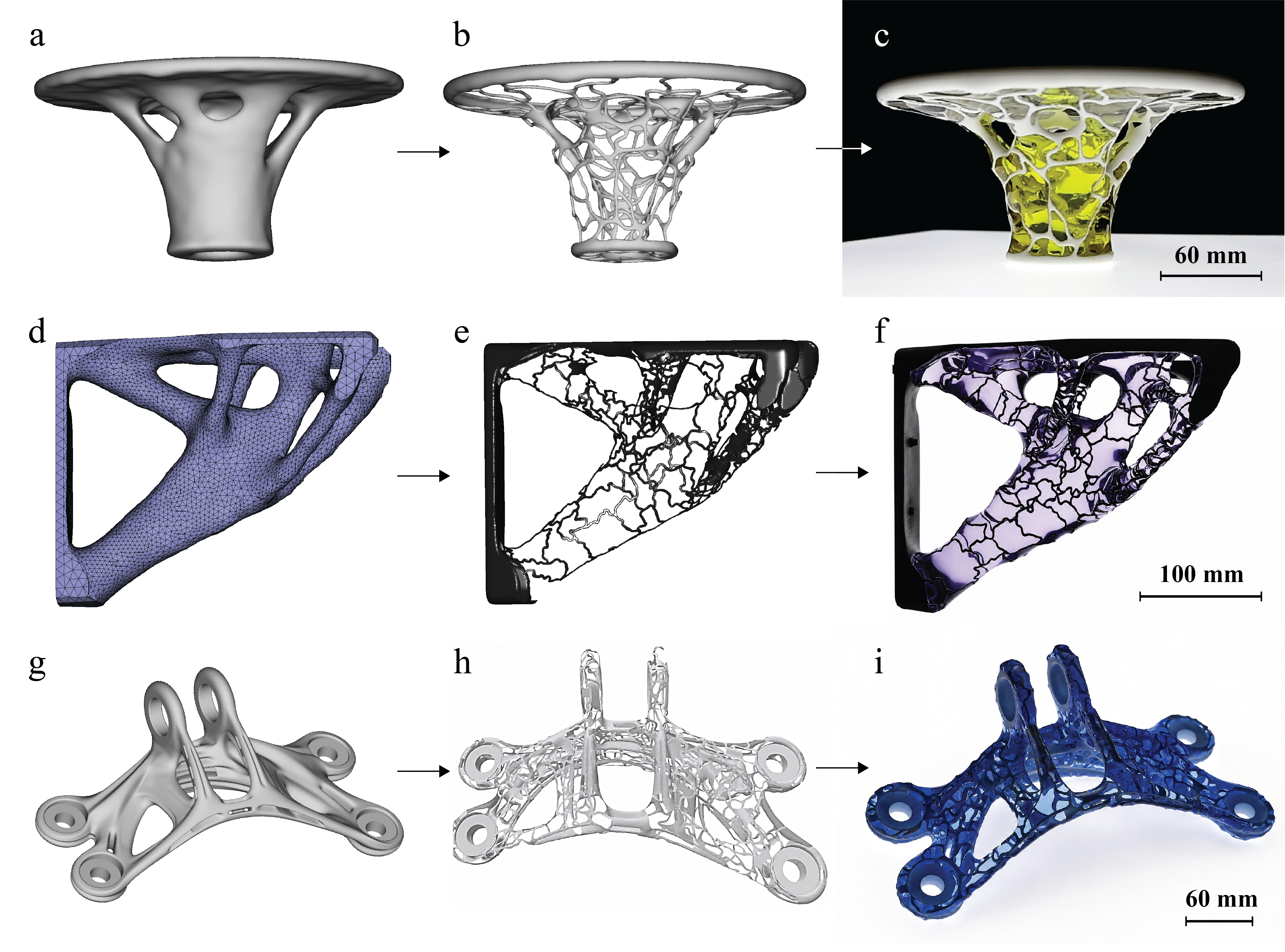}
\par\end{centering}
    \caption{\small{Topology optimized designs fabricated using Fluidic Shaping. (a-c) 3D model designed using nTop 5.45.2 \cite{noauthor_ntop_2026}, wireframe segmentation, and final fabrication of a TO platform. The wireframe is made of PLA (Bambu Labs), while the stem is made of Vida Rosa TJ-3704A and the platform surface is made of Elastosil RT601 (Wacker). (d-f) 3D model obtained from MakerWorld \cite{janosiano_organic_2025}, wireframe segmentation, and final fabrication of a TO wall-mounted bracket. The wireframe and mounting points are made of PLA (Bambu Labs), while the structural support is made of Vida Rosa TJ-3704A. (g-i) 3D model obtained from GrabCAD \cite{anedda_ge_2013}, wireframe segmentation, and final fabrication of a bracket submitted for the General Electric jet engine bracket challenge \cite{noauthor_ge_nodate}. The wireframe is made of PLA (Bambu Labs), and the bulk structure is made of Vida Rosa TJ-3704A.}}
    \label{fig:3}
\end{figure}

We proceed by demonstrating how interfacial self-organization can be used for the fabrication of macroscopic TO designs with complex geometric features. Figure~\ref{fig:3} depicts three examples, showing the original 3D design, the extracted wireframe, and the final fabricated part. Figure~\ref{fig:3}a–c shows the wireframe decomposition and fabrication of a TO platform generated using nTop 5.45.2 (nTop, New York City, NY, USA) \cite{noauthor_ntop_2026}, derived from a simple frustum with a surface load applied to the top surface (see Supplementary Information). The native TO algorithm generated complex branching geometry beneath the top surface and a cavity within the stem. The wireframe in Fig.~\ref{fig:3}c was 3D printed using PLA, the stem was \emph{Fluidically Shaped} using a UV curable resin (Vida Rosa TJ-3704A), while the platform surface was \emph{Fluidically Shaped} with a soft elastomer (Elastosil RT601). This example illustrates the ability of {\it Fluidic Shaping} to incorporate several mechanically different materials into a single structure. The TO brackets shown in Fig.~\ref{fig:3}d–i are two designs sourced from MakerWorld \cite{janosiano_organic_2025} and GrabCAD \cite{anedda_ge_2013}, respectively, each exhibiting similarly intricate geometric features such as internal cavities and highly curved surfaces. The wireframes for these parts were 3D printed using PLA, where the \emph{Fluidic Shaping} material used was Vida Rosa TJ-3704A. In all cases, parts were fabricated directly from their TO outputs, with no additional manufacturing constraints for \emph{Fluidic Shaping}, yielding components more faithful to the optimized design and free from the performance penalties commonly associated with constraint-modified TO procedures. 

\subsection*{Discussion}

The results presented here establish a new manufacturing approach in which sparse geometric scaffolds can be transformed into complex three-dimensional structures through interfacial self-organization. Unlike conventional manufacturing approaches, which directly prescribe the geometry of a structure through material deposition, removal, or molding, \textit{Fluidic Shaping} translates the target geometry into a set of physically realizable CMC domains and their shared boundaries. The non-uniqueness of this decomposition allows one to balance geometric fidelity against wireframe complexity, exposed smooth-surface area, and structural reinforcement, for the same target.
Although the smooth-surface is interrupted by the supporting wireframe, it constitutes nearly the entire exposed surface area, a fraction that can be further increased through wireframe optimization, e.g., by reducing wireframe compactness.

Because the bulk material is introduced as a continuous liquid and subsequently solidified, {\it Fluidic Shaping} is not intrinsically tied to the layerwise construction characteristic of most additive processes. This allows fabrication of mechanically isotropic bulk materials while retaining complex three-dimensional geometry and the intrinsic smoothness of liquid interfaces. 
The method is compatible with a broad range of polymers, elastomers, thermosets, and other curable materials (Materials and Methods), and different materials can be integrated within a single structure. The wireframe can remain embedded in the final object, allowing it to serve both as a geometric constraint during shaping and as structural reinforcement or a second functional material after solidification.

A current bottleneck of the method is the fabrication of the wireframe itself. In this work, the wireframes were produced by conventional 3D printing, which limits the fabrication speed and scale of the overall process. However, because the wireframe consists primarily of one-dimensional curves, it may be fabricated using substantially faster and more scalable methods, including automated wire bending \cite{bhundiya_bend-forming_2023} or extrusion \cite{kringer_direct_2022}. For geometries that can be constructed from planar elements, laser cutting combined with folding or kirigami \cite{lamoureux_kirigami-inspired_2025, jin_kirigami-inspired_2020} provides another possible route. Developing such approaches will be important for extending {\it Fluidic Shaping} to larger structures.

\subsubsection*{Competing interests:}
V.F. is an inventor on a provisional patent application (U.S. Provisional Patent Appl. No. 64/101,845) filed by Boston University on June 30, 2026, covering aspects of the manufacturing method described in this manuscript. The other authors declare no competing interests.

\subsection*{Funding:}
E. C. acknowledges support from the NSF CAREER award (2543971). E. W. aknowledges support from the NSF CAREER award (2047342).

\bibliographystyle{unsrt}
\bibliography{FSCG}

\end{document}